\documentclass[12pt,english]{article}
\pdfoutput=1
\usepackage[T1]{fontenc}
\usepackage[utf8]{inputenc}
\usepackage{subfigure,lmodern, amsmath,amssymb, graphicx, pifont, adjustbox, bm, xcolor}
\usepackage{amsfonts}
\usepackage{geometry}
\usepackage{comment}
\usepackage{mathtools}
\usepackage{float}
\usepackage{slashed}
\usepackage{cite}
\usepackage{ragged2e}
\usepackage{etoolbox}
\usepackage{array}
\usepackage{setspace}
\usepackage[hang,flushmargin]{footmisc}

\apptocmd{\thebibliography}{\justifying\setlength{\leftskip}{7.4mm}}{}{}
\makeatletter\g@addto@macro\bfseries{\boldmath}\makeatother
  
\usepackage{stackengine}
\usepackage{esint}
\usepackage[unicode=true,pdfusetitle,
 bookmarks=true,bookmarksnumbered=false,bookmarksopen=false,
 breaklinks=false,pdfborder={0 0 1},backref=false,colorlinks=true]
 {hyperref}
\hypersetup{pdfauthor={Ning Bao, Grant N. Remmen},
 citecolor=black,linkcolor=black,urlcolor=black}

\usepackage{changepage}

\newcommand{\be}{\begin{equation}}
\newcommand{\ee}{\end{equation}}
\newcommand{\bea}{\begin{eqnarray}}
\newcommand{\eea}{\end{eqnarray}}

\begin{document}

\interfootnotelinepenalty=10000
\widowpenalty10000
\baselineskip=24pt
\hfill 
\hfill
\vspace{1.5cm}
\thispagestyle{empty}
\begin{center}
{\LARGE \bf
Black Hole Complementarity and ER/EPR
}\\
\bigskip
\bigskip
\begin{center}{\large Ning Bao${}^{a,b}$ and Grant N. Remmen${}^{c}$}\end{center}
{
\it  ${}^a$Department of Physics, Northeastern University, Boston, MA 02115 \\
${}^b$Computational Science Initiative, Brookhaven National Laboratory, Upton, NY 11973\\
${}^c$Center for Cosmology and Particle Physics, Department of Physics,\\[-2.3mm] New York University, New York, NY 10003}
\let\thefootnote\relax\footnote{e-mail:  
\url{ningbao75@gmail.com}, 
\url{grant.remmen@nyu.edu}  }
\end{center}

\bigskip
\centerline{\large\bf Abstract}
\begin{quote} \small \doublespacing
We demonstrate that wormholes must be entangled regardless of asymptotic boundary conditions. Assuming black hole complementarity, we argue that traversable wormholes instantiate entanglement-assisted quantum channels and that this entanglement must be present between the stretched horizons as an initial condition prior to traversability. This result demonstrates the forward direction of the ER/EPR conjectures.
\vspace{1mm}

\scalebox{0.95}[1]{\it Honorable Mention, Gravity Research Foundation 2025 Awards for Essays on Gravitation}

%\hfill {\it March 20, 2025}

\end{quote}
	
\setcounter{footnote}{0}

\newpage

\setcounter{page}{1}
\section{Introduction}
Wormholes and quantum entanglement share the qualitative characteristic of connecting otherwise distinct degrees of freedom at great distance.
In the anti-de Sitter/conformal field theory (AdS/CFT) correspondence, the fact that wormholes are equivalent to a specific pattern of entanglement, specifically the thermofield double state,
\begin{equation}
|\psi\rangle = \sum_n e^{-\beta E_n/2} |\bar n\rangle_L\otimes|n\rangle_R,
\end{equation}
in the dual boundary theory is well established~\cite{Maldacena:2001kr,van2010building}, where $\beta^{-1}$ is the temperature, $Z$ the partition function, $|n\rangle$ the energy eigenstates, and the bar denotes CPT conjugation.
The ER/EPR proposal~\cite{Maldacena_2013} goes further, suggesting a formal equivalence of entanglement between quantum systems and wormholes connecting these quantum systems, in other words, a mapping between Einstein-Rosen bridges and Einstein-Podolsky-Rosen pairs.
Beyond the connection between wormholes and entanglement in AdS/CFT, this claim posits that the relationship {\it i}.~is bijective, {\it ii}.~is true regardless of the asymptotic boundary conditions of the spacetime, and {\it iii}.~remains true even for entanglement between small numbers of qubits, with the wormhole in this case proposed to be a quantum gravitational or stringy wormhole. 

Aspects of the ER/EPR correspondence have been verified in Refs.~\cite{Bao_2015_1,Bao_2015_2,Remmen_2016}, where in the limit of semiclassical wormholes, gravitational analogues were derived for quantum properties such as the no-cloning theorem, the undetectability of entanglement via a linear operator (i.e., as a Dirac observable), and conservation of entanglement across a bipartition via local operations and classical communication (LOCC). There have also been attempts to understand the stringy limit of the ER/EPR proposal, as in Ref.~\cite{Jafferis_2022}, and explorations of black hole entropy and its connections to holographic wormholes and fine-grained entanglement structure in Refs.~\cite{Engelhardt:2017aux,Engelhardt:2018kcs,Nomura:2018aus,Bousso:2018fou,Bao:2018fso,Freedman:2016zud,Bao:2019wcf,Bao:2017thr,Bao:2021vyq,Bao:2021ebo,Chatwin-Davies:2023ofu}. Furthermore, it has been shown that not every pattern of entanglement corresponds to semiclassical wormholes, particularly in the context of AdS/CFT, as certain entangled states would violate the holographic entanglement entropy inequalities~\cite{Hayden_2013,Bao_2015} enforced by the Ryu-Takayanagi formula~\cite{Ryu_2006,Hubeny:2007xt}. 

From a causal perspective, a nontraversable wormhole, such as that formed by the maximal analytic extension of the Schwarzschild geometry, need not exhibit entanglement between the two exterior regions.
Nonetheless, there is an intuitive sense in which we expect wormholes to be entangled, even outside of the context of AdS/CFT.
For example, pair production via a gravitational instanton in an electromagnetic field produces highly entangled extremal Reissner-Nordstr\"om black holes~\cite{Garfinkle:1990eq,Garfinkle:1993xk}.
Whether this feature holds for all pairs of black hole horizons with a common interior, in spacetimes with arbitrary asymptotics, remains an open question~\cite{forthcoming}.

It is this problem that we will address in the present work. In particular, we will make use of tools from both relativity and quantum information---specifically, traversable wormhole constructions and quantum channels---to show that {\it all} wormholes in a common exterior spacetime must share entanglement between their exterior degrees of freedom, regardless of asymptotic boundary conditions, provided that black hole complementarity is correct~\cite{Susskind_1993}.
While the true entanglement structure of horizons is itself an extremely difficult problem in quantum gravity, as sharpened by the firewall formulation of the black hole information paradox~\cite{Almheiri_2013,Bao:2017who,Bousso:2012as}, it will be striking that, crucially provided that complementarity holds in a sense that we will make clear, the general conclusion that horizons with a shared interior must be entangled follows from a weaker set of assumptions than AdS/CFT alone.

The organization of this essay is as follows. Taking a self-contained approach, in Sec.~\ref{sec:BHC} we review the relevant details of traversable wormholes and black hole complementarity. In Sec.~\ref{sec:entanglement}, we combine these concepts to show that black hole complementarity implies entanglement of the stretched horizons of traversable wormholes. 
We argue in Sec.~\ref{sec:RS} that this entanglement cannot be generated by the process that renders the wormhole traversable, but must be present as an initial condition for the stretched horizons of any wormhole. We make concluding remarks in Sec.~\ref{sec:discussion}.

\newpage

\section{Wormholes and Black Hole Complementarity}\label{sec:BHC}
\subsection{Review of Traversable Wormholes}\label{sec:traversable}
In our study of wormholes and their associated entanglement, we will make use of the tool of {\it traversable} wormholes. 
Traversable wormholes were first proposed in general relativity independently in Refs.~\cite{Ellis:1973yv,Bronnikov:1973fh,10.1119/1.15620} and were studied in the context of the AdS/CFT correspondence in Refs.~\cite{Gao_2017, Maldacena_2017}.
As we will discuss in Sec.~\ref{sec:entanglement}, black hole complementarity offers an interpretation of these constructions in terms of entanglement-assisted quantum channels.

\begin{figure}[t]
    \centering
    \includegraphics[width=9cm]{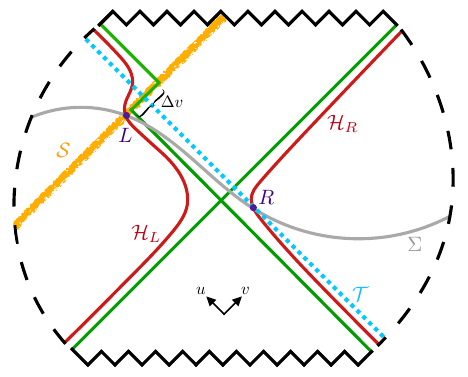}
    \caption{\onehalfspacing Penrose diagram for the wormhole quantum channel protocol, in a spacetime with arbitrary asymptotics (dashed lines). A NEC-violating shock wave ${\cal S}$ (yellow) drags the left apparent horizon (green) by $\Delta v$. This makes the wormhole traversable by a signal ${\cal T}$ (blue) sent from the right side. In the black hole complementarity picture, this process can be viewed as an entanglement-assisted quantum channel between the left and right stretched horizons ${\cal H}_{L,R}$ (red). In particular, we show that there must exist entanglement between the surfaces $L$ and $R$ (purple) at which ${\cal S}$ and ${\cal T}$ enter their respective stretched horizons; $L$ and $R$ both live in some Cauchy slice $\Sigma$ (gray).\vspace{5mm}\label{fig:wormhole}} 
\end{figure}

The way in which a nontraversable wormhole can be rendered traversable is via the insertion of negative stress-energy shock waves into the horizons. Considering the rotating BTZ black hole in $D=3$ spacetime dimensions as an example~\cite{Banados:1992wn} (though the construction will be similar in arbitrary $D$), let us work in Kruskal-like null coordinates $u$ and $v$ such that, near the horizon, the metric becomes
\begin{equation}
{\rm d}s^2 \rightarrow -4\,{\rm d}u{\,\rm d}v + r_{\rm h}^2 {\rm d}\phi^2,
\end{equation}
where the horizons are located at $uv=0$.
As discussed in Refs.~\cite{Shenker:2013pqa,Bao_2018}, 
if we insert a shock wave with $T_{uu}=-\alpha\delta(u-u_0)/4\pi G$, the near-horizon geometry shifts via the replacement $v\rightarrow \hat v = v-\alpha\theta(u-u_0)$, to 
\begin{equation}
{\rm d}s^2 = -4\,{\rm d} u \,{\rm d}\hat v + 4\alpha\,\delta(u-u_0){\rm d}u^2 + r_{\rm h}^2 {\rm d}\phi^2.
\end{equation}
The net effect is that the left-hand horizon acquires a Shapiro time advance $\Delta v=\alpha$ at $u=u_0$; see Fig.~\ref{fig:wormhole}.
This allows a particle starting from one asymptotic boundary to reach the other asymptotic boundary without encountering the singularity.

In the context of holography, these shock waves are interpreted as double trace deformations on the boundary CFTs. 
Consider a bundle of null geodesics---with affine parameter $\lambda$, tangent $k^\mu=({\rm d}/{\rm d}\lambda)^\mu$, expansion $\theta = \nabla_\mu k^\mu$, and shear $\sigma_{\mu\nu}$---passing through the wormhole. Prior to entering the wormhole, $\theta<0$, while $\theta>0$ upon exit. The Raychaudhuri equation,
\begin{equation}
\frac{{\rm d}\theta}{{\rm d}\lambda} = -\frac{1}{D-2}\theta^2 - \sigma_{\mu\nu}\sigma^{\mu\nu} - R_{\mu\nu}k^\mu k^\nu,
\end{equation}
and the Einstein equation together therefore imply that the null energy condition (NEC) $T_{\mu\nu}k^\mu k^\nu \geq 0$ must be violated for the shock wave, a famous pathology of traversable wormholes.
A stronger statement is the averaged null energy condition (ANEC), which requires
\begin{equation}
\int_{-\infty}^{\infty} T_{\mu\nu} k^\mu k^\nu {\rm d}\lambda \geq 0,\label{eq:ANEC}
\end{equation}
integrated along an inextensible null geodesic.
A traversable wormhole will necessarily violate the ANEC as well, but for quantum fields in curved spacetime it is only the achronal ANEC---that is, the ANEC restricted to null geodesics on which no two points can be connected by a timelike trajectory---that is expected to apply~\cite{Graham:2007va,Wald:1991xn,Visser:1994jb,Urban:2009yt}. The recent proof of the ANEC~\cite{Hartman:2023qdn} contemplates a non-affinely-parameterized quantity in place of Eq.~\eqref{eq:ANEC}, and the ANEC in Eq.~\eqref{eq:ANEC} can indeed be violated for chronal null geodesics.
In physical traversable wormhole constructions, ANEC-violating null geodesics are indeed chronal. The double-trace deformation~\cite{Gao_2017, Maldacena_2017}, for example, puts the two boundaries in causal contact, while the construction of Ref.~\cite{Maldacena:2018gjk} based on Casimir energy builds ``long wormholes,'' in which the two wormhole mouths are closer together in the exterior spacetime than in the wormhole throat.
In such cases, violation of the ANEC is permissible.
For any wormhole in which the two mouths are in the same asymptotic spacetime, we can imagine rendering it traversable by, for example, gently gravitationally towing the two mouths near to each other and then implementing one of the wormhole traversability protocols, with the null shock wave entering at sufficiently large $u_0$ that the wormhole is indeed long, preserving the achronal ANEC.

\subsection{Review of Black Hole Complementarity}
Black hole complementarity~\cite{Susskind_1993,tHooft:1984kcu} is the idea that the information associated with a black hole can equivalently be thought of as living in two places: either at the stretched horizon on the black hole~\cite{MembraneParadigm}, where it remains accessible to external observers, or in the black hole interior, which is only accessible by observers who have crossed the event horizon.
The stretched horizon is the timelike worldvolume of the spacelike surface at a fixed cutoff (e.g., a Planck length) from the horizon. In the effective field theory of the external observer, degrees of freedom between the stretched horizon and the event horizon are inaccessible.
For concreteness, as shown in Fig.~\ref{fig:wormhole}, we label the stretched horizon on the left or right side of the wormhole ${\cal H}_L$ and ${\cal H}_R$, respectively; we will label our shock wave ${\cal S}$ and our transmission signal through the wormhole ${\cal T}$.
The apparent violation of the no-cloning theorem in black hole complementarity is avoided by ensuring that no observer simultaneously has access to both copies of the information associated with the black hole.

Black hole complementarity was a primary motivator for the formulation of the AdS/CFT correspondence. In particular, the holographic description of the black hole interior encoded onto the stretched horizon was a direct precursor to the holographic description of the asymptotically AdS gravitational spacetime encoded in the boundary CFT~\cite{Susskind_1993,SusskindHolography,tHooft:1984kcu,tHooft:1993dmi,Bousso:2002ju,AdSCFT,MAGOO,Witten:1998qj,SusskindWitten}. As such, many analogies can be drawn between AdS/CFT and black hole complementarity, a fact that we will leverage later in this essay.

Complementarity has come under criticism over the past fifteen years in the context of the firewall paradox~\cite{Almheiri_2013,Bao:2017who,Bousso:2012as}. 
While some in the community have proposed that entanglement islands~\cite{penington2020entanglementwedgereconstructioninformation,Almheiri_2019} can resolve aspects of the information paradox in a way that has been suggested to preserve black hole complementarity, consensus has not yet been reached~\cite{Geng:2020qvw,Geng:2020fxl,Geng:2021hlu,Raju:2021lwh,Geng:2021wcq,Bousso:2023kdj,Antonini:2025sur} as to whether islands fully solve the information paradox in its modern firewalls form or whether complementarity itself fully survives.

\section{Complementarity Implies Entangled Wormholes}\label{sec:entanglement}

A quantum channel is a completely general positive trace preserving map, and for systems in which a trace exists it always can be written as the partial trace over some unitary matrix,
\begin{equation}
\varepsilon (\rho_A)={\rm Tr}_{B} U_{AB}.
\end{equation}
A quantum communication channel is a channel that is used to send information from one location to another, and an entanglement-assisted quantum communication channel is one that is instantiated by pre-existing entanglement between the two locations, as in the quantum teleportation protocol.\footnote{We note in passing that quantum channels have also recently been of interest in the study of noninvertible symmetries, such as in Ref.~\cite{Okada_2024}; we leave the interesting question of whether noninvertible symmetries can cast light on the formulation of wormholes as quantum channels~\cite{Bao_2018} to future work.}
In the holographic context, it was proposed in Ref.~\cite{Gao_2017} that traversable wormholes are dual to quantum communication channels between the causally disconnected CFTs on the boundaries, connected by the wormhole through the bulk. The definition of this object as a quantum communication channel, specifically an entanglement-assisted quantum communication channel, was made more precise in Ref.~\cite{Bao_2018}, in particular with respect to its channel capacity properties.

The fact that LOCC between observers outside of a traversable wormhole is equivalent to a quantum channel in the complementary frame was established by Gao, Jafferis, and Wall in the context of AdS/CFT~\cite{Gao_2017}. In our context, we will not be using AdS/CFT but instead the broader framework of black hole complementarity. As we have emphasized, in black hole complementarity all processes behind the horizon correspond to some process on the quantum state defined on the stretched horizons. The most general process on any quantum state is a quantum channel, so without loss of generality, we can view the transmission of information through a traversable wormhole, in the black hole complementarity picture, as some quantum channel acting on the two stretched horizons. As should be clear, this interpretation hinges crucially on the validity of black hole complementarity, so if the final resolution of the information paradox turns out to violate complementarity, our conclusions will not hold.
In particular, the consumption of entanglement in an entanglement assisted quantum channel always relies on a measurement occurring~\cite{bennett2002entanglementassistedcapacityquantumchannel}. While the explicit form of such a measurement in the stretched horizon picture is not immediately obvious, in the complementary frame of the black hole interior, the negative null energy shock wave plays this role by definition---as explicitly explored  in Ref.~\cite{Gao_2017}---since the shock wave is the only object in the interior that is not the message itself. We leave a detailed exploration of the measurement problem and decoherence on the stretched horizon side of the duality in black hole complentarity to future work.

Importantly, the argument that a traversable wormhole has an interpretation as a quantum channel did not require AdS/CFT. In fact, the negative energy shock waves in the original proposal for traversable wormholes in Ref.~\cite{10.1119/1.15620} did not require asymptotically AdS boundary conditions at all. The connection to AdS/CFT gave a nice motivation for considering such negative energy shock waves as double trace deformations in the CFT, but this is not necessary for the creation of a quantum channel. So long as a source of negative energy shock waves can be produced in the bulk, e.g., via the Casimir construction of Ref.~\cite{Maldacena:2018gjk}, a quantum communication channel can be established.

From the perspective of someone traversing the wormhole, the transmission of information from one side to the other is unsurprising: qubits are physically sent through the wormhole via the transmission ${\cal T}$.
The statement becomes nontrivial, however, when one considers the perspective of two observers who remain outside of the wormhole region and retain access to only the information encoded on their respective stretched horizons ${\cal H}_L$ and ${\cal H}_R$. 
We can restrict communication within the exterior spacetime between the two external observers to LOCC.
By black hole complementarity, any quantum information associated with the qubits sent via ${\cal T}$ through the wormhole disappears from one stretched horizon and reappears on the other. As such, from the perspective of outside observers and the stretched horizons, the traversable wormhole implemented a quantum communication channel that was not possible using purely classical means of communication in the exterior.

Entanglement-assisted quantum channels are the only channels capable of transmitting quantum information between two regions without directly transporting the qubit through the ambient spacetime~\cite{bennett2002entanglementassistedcapacityquantumchannel,May:2019yxi}, which we have forbidden by restricting external communication between the two observers to LOCC. The amount of entanglement that is consumed per qubit transmitted in an entanglement-assisted quantum channel will in general depend on the protocol; for example, in the quantum teleportation channel, one Bell pair's worth of entanglement must be consumed in order to send one qubit's worth of quantum information.
The properties of such channels in the context of the AdS/CFT traversable wormhole were detailed in Ref.~\cite{Bao_2018}.
The presence of an entanglement-assisted quantum channel necessitates entanglement between the two stretched horizons, i.e., the traversable wormhole is entangled. 

We can lower-bound the necessary entanglement present in the wormhole as follows.
The one-qubit channel capacity $C_1(\varepsilon)$---that is, the maximal amount of information in qubits conveyed by the channel $\varepsilon$ when one qubit is sent through from a system $L$ to $R$---is upper-bounded by $1$. 
More broadly, in entanglement-assisted quantum channels, the channel capacity for an $n$-qubit transmission is upper-bounded by $n$, which is by definition upper-bounded by the number of Bell pairs $N$ distillable between $L$ and $R$. Finally, writing the von~Neumann entropy $S(A)=-{\rm Tr}\rho_A \log \rho_A$, we have that the mutual information between subsystems,
\begin{equation}
I(L:R)=S(L)+S(R)-S(LR),
\end{equation}
constitutes an upper bound to the distillable entanglement~\cite{bennett2002entanglementassistedcapacityquantumchannel}, as it in particular could also contain non-distillable Shannon contributions.
 All together,
\begin{equation}
C_n(\varepsilon)\leq n \leq N \leq I(L:R),
\end{equation}
so since our traversable wormhole achieves some nonzero transmission of information via our signal ${\cal T}$, from the black hole complementarity perspective of the exterior observer, there must be nonzero mutual information between the two mouths $L$ and $R$.\footnote{In particular, we expect that the maximal amount of information transmitted---and hence our expected lower bound on the entanglement between the horizons---scales like some small parameter $\delta \ll 1$ times ${\cal A}/4G$ for ${\cal A}$ of order the area of the horizons, since we can transmit $\delta\times O({\cal A}/4G)$ qubits while still remaining in perturbative control of the geometry. We leave further quantification of the horizon entanglement to future work.}
We emphasize that this conclusion is reached without appeal to AdS/CFT, but directly as a consequence of black hole complementarity in a spacetime with arbitrary asymptotics.
While it is possible to consider quantum channels associating more than two subsystems, for this work it is sufficient to restrict to two subsytems, and thus the number of distillable Bell pairs is sufficient to characterize the entanglement, rather than more general forms of squashed entanglement.

\section{Entangled Horizons as an Initial Condition}\label{sec:RS}

We have shown that, once rendered traversable,\footnote{Recall that in Ref.~\cite{Maldacena:2018gjk}, the protocol for generating the NEC-violating shock wave that makes the wormhole traversable is through Casimir effects when the two horizons are sufficiently close together; one generates a NEC-violating shock wave by simply taking two horizons with a shared interior and gravitationally towing them to be close together in the exterior spacetime. In Ref.~\cite{Gao_2017}, the shock wave protocol is instead the turning on of the double trace deformation coupling in the AdS/CFT dual.} wormholes must have stretched horizons that are entangled with each other.
We will now argue that this entanglement must be {\it intrinsic} to the wormhole itself---rather a byproduct of traversability---so that the entanglement predated the null shock wave.
That is, all two-sided wormholes, traversable or not, must be entangled if the two sides are in the same asymptotic spacetime.

First, we can eliminate the possibility that the shock wave itself carried the entanglement. 
Nothing in general relativity required entanglement to engineer a time advance; the NEC-violating shock wave can be entirely decohered, and we can regard it as a LOCC process.

For the wormholes we consider, the two stretched horizons must necessarily be in causal contact, in order for us to be permitted to make the wormhole traversable without violating the achronal ANEC as discussed in Sec.~\ref{sec:traversable}. While causally connected regions of space in quantum field theory are always entangled in the vacuum, by virtue of the Reeh--Schlieder theorem~\cite{cmp/1103758945}, this conclusion does not apply to  entanglement of the wormhole mouths for two reasons. First, the state in the exterior need not necessarily be the vacuum; further, and more crucially, the Reeh--Schlieder theorem applies to local relativistic quantum field theories, i.e., to Type~III von~Neumann algebras. We note that the von~Neumann algebra relevant to stretched horizons of black holes is Type~I, as the number of degrees of freedom is finite due to the finiteness of Bekenstein-Hawking entropy~\cite{Bekenstein:1973ur,HawkingParticle}.
A modern take on this problem by Witten et al.~\cite{strohmaier2023timeliketubetheoremcurved} in the context of the timelike tube theorem similarly shows that the von~Neumann algebra type changes once coupled to gravity~\cite{Chandrasekaran_2023}. 
Even if some weaker version of Reeh--Schlieder applied morally in the wormhole setup, one still could not prove that the entanglement implied would be usable for any quantum information theoretic task; indeed, it may be analogous to the unusable entanglement of the fermionic vacuum~\cite{Zanardi_2004}.
Indeed, any argument that would imply that Reeh--Schlieder-like entanglement in the ambient spacetime surrounding each observer is responsible for the entanglement assisted quantum channel that enables their communication would equally well apply to two horizons that do not have a shared interior, and communication between such horizons---in the absence of LOCC through the exterior spacetime, which we have forbidden---is not possible. Moreover, we can always idealize our observer as a finite-dimensional quantum system, simply by considering a quantum computer with a finite number of qubits coupled to the stretched horizon. While we could consider a field-theoretic observer endowed with a Type~III algebra, it suffices to demonstrate the necessity of entanglement between the two horizons in the case of finite-dimensional observers.

A final possibility is that the back reaction on the wormhole region from the classical shock wave is enough to permit entanglement between the two asymptotic regions, via quantum gravitational degrees of freedom propagating through the wormhole once it is rendered traversable.
Indeed, it is clear from Fig.~\ref{fig:wormhole} that there always exists some spacelike Cauchy surface $\Sigma$ containing both $L$ and $R$, where $L = {\cal S}\cap {\cal H}_L$, the intersection of the shock wave with the left stretched horizon of the wormhole, and $R = {\cal T}\cap {\cal H}_R$, the entry of our signal into the right stretched horizon, so $L$ and $R$ cannot be unambiguously causally ordered.
However, it is possible to make the proper distance between any point in $L$ and any point in $R$ arbitrarily large, by simply sending in the NEC-violating shock wave at some very large value of $u_0$. (Interestingly, taking $u_0$ large was also necessary in order to guarantee compliance with the achronal ANEC, as we noted in Sec.~\ref{sec:traversable}.)
In order for the entanglement to {\it not} be intrinsic to the wormhole (i.e., to not predate the shock wave's arrival), and thus for quantum gravitational back reaction to the shock wave to be responsible for the entanglement between the two stretched horizons, the entanglement would be required to travel arbitrarily close to the speed of light.
However, the Lieb-Robinson bound---as applied to the local system seen by the observer traveling through the wormhole---precludes the entanglement from propagating fast enough.
Specifically, given two observables ${\cal O}_L$ and ${\cal O}_R$ defined on the finite surfaces $L$ and $R$, if we consider a time-dependent excitation ${\cal O}_L(t)$, there exist some positive constants $\lambda,c$ and a Lieb-Robinson velocity $v_{LR}\leq 1$ for which
\begin{equation}
    || [{\cal O}_L(t),{\cal O}_R]||\leq c\, e^{-\lambda (d(L,R) -v_{LR}|t|)},
\end{equation}
where $d(L,R)$ is the proper distance and $||\cdot||$ is the operator norm~\cite{cmp/1103858407}.
Estimates of the butterfly velocity $v_B$---essentially an effective Lieb-Robinson velocity for chaotic systems---have been made~\cite{Roberts_2016,Roberts_2014}, and in large-$N$ holographic systems one has $v_B = \sqrt{\frac{D+1}{2D}}$.
While we will not use the precise holographic result, as long as $v_{LR} < 1$, quantum gravitational back reaction is precluded from being able to generate the entanglement at the requisite separation.
The only remaining option is that the stretched horizons of the wormhole were entangled before the traversability process began, simply by virtue of having a shared interior. 

\section{Discussion}\label{sec:discussion}
In this work, we have shown that any two horizons that share an external spacetime and interior must be entangled, even if not traversable, in spacetimes with arbitrary asymptotics.
We do not comment on more detailed aspects of the structure of the reduced density matrix on the stretched horizon, except that the entanglement must be such that an entanglement-assisted quantum channel can be constructed. It is in principle possible that there exist specific patterns of, e.g., multipartite entanglement that do not allow for entanglement-assisted channels, but we will leave the exploration of this possibility to future work. 

If one takes the restrictions of types of entanglement---i.e., the requirements on the fine-grained structures of permitted entangled states---from AdS/CFT as applying to more general spacetimes, then this would immediately rule out certain patterns of entanglement, such as GHZ states, even though GHZ entanglement can be used as a resource in an entanglement-assisted quantum channel. This would suggest that not every pattern of entanglement that enables a quantum communication channel has an associated semiclassical wormhole, so that EPR does not imply ER; conversely, our results show that, with black hole complementarity, ER indeed implies EPR.

\vspace{5mm}

\section*{Acknowledgments}
We thank Aidan Chatwin-Davies, Ahmed Almheiri, and Aidan Herderschee for useful conversations. N.B. is supported by Northeastern University Department of Physics and by Brookhaven National Laboratory. 
G.N.R. is supported by the James Arthur Postdoctoral Fellowship at New York University.

\pagebreak

\singlespacing
\bibliographystyle{utphys-modified}
\bibliography{wormholes}

\providecommand{\href}[2]{#2}\begingroup\raggedright\begin{thebibliography}{10}

\bibitem{Maldacena:2001kr}
J.~M. Maldacena, ``{Eternal Black Holes in Anti-de~Sitter},''
  \href{http://dx.doi.org/10.1088/1126-6708/2003/04/021}{{\em JHEP} {\bfseries
  0304} (2003) 021},
\href{http://arxiv.org/abs/hep-th/0106112}{{\ttfamily arXiv:hep-th/0106112
  [hep-th]}}.
%%CITATION = HEP-TH/0106112;%%.

\bibitem{van2010building}
M.~Van~Raamsdonk, ``{Building Up Spacetime with Quantum Entanglement},''
  \href{http://dx.doi.org/10.1142/S0218271810018529}{{\em Gen. Rel. Grav.}
  {\bfseries 42} (2010) 2323}, \href{http://arxiv.org/abs/1005.3035}{{\ttfamily
  arXiv:1005.3035 [hep-th]}}.

\bibitem{Maldacena_2013}
J.~Maldacena and L.~Susskind, ``{Cool Horizons for Entangled Black Holes},''
  \href{http://dx.doi.org/10.1002/prop.201300020}{{\em Fortsch. Phys.}
  {\bfseries 61} (2013) 781}, \href{http://arxiv.org/abs/1306.0533}{{\ttfamily
  arXiv:1306.0533 [hep-th]}}.

\bibitem{Bao_2015_1}
N.~Bao, J.~Pollack, and G.~N. Remmen, ``{Splitting Spacetime and Cloning
  Qubits: Linking No-Go Theorems across the ER=EPR Duality},''
  \href{http://dx.doi.org/10.1002/prop.201500053}{{\em Fortsch. Phys.}
  {\bfseries 63} (2015) 705--710},
  \href{http://arxiv.org/abs/1506.08203}{{\ttfamily arXiv:1506.08203
  [hep-th]}}.

\bibitem{Bao_2015_2}
N.~Bao, J.~Pollack, and G.~N. Remmen, ``{Wormhole and Entanglement
  (Non-)Detection in the ER=EPR Correspondence},''
  \href{http://dx.doi.org/10.1007/JHEP11(2015)126}{{\em JHEP} {\bfseries 11}
  (2015) 126}, \href{http://arxiv.org/abs/1509.05426}{{\ttfamily
  arXiv:1509.05426 [hep-th]}}.

\bibitem{Remmen_2016}
G.~N. Remmen, N.~Bao, and J.~Pollack, ``{Entanglement Conservation, ER=EPR, and
  a New Classical Area Theorem for Wormholes},''
  \href{http://dx.doi.org/10.1007/JHEP07(2016)048}{{\em JHEP} {\bfseries 07}
  (2016) 048}, \href{http://arxiv.org/abs/1604.08217}{{\ttfamily
  arXiv:1604.08217 [hep-th]}}.

\bibitem{Jafferis_2022}
D.~L. Jafferis and E.~Schneider, ``{Stringy ${\rm ER}\,{=}\,{\rm EPR}$},''
  \href{http://dx.doi.org/10.1007/JHEP10(2022)195}{{\em JHEP} {\bfseries 10}
  (2022) 195}, \href{http://arxiv.org/abs/2104.07233}{{\ttfamily
  arXiv:2104.07233 [hep-th]}}.

\bibitem{Engelhardt:2017aux}
N.~Engelhardt and A.~C. Wall, ``{Decoding the Apparent Horizon: Coarse-Grained
  Holographic Entropy},''
  \href{http://dx.doi.org/10.1103/PhysRevLett.121.211301}{{\em Phys. Rev.
  Lett.} {\bfseries 121} (2018) 211301},
  \href{http://arxiv.org/abs/1706.02038}{{\ttfamily arXiv:1706.02038
  [hep-th]}}.

\bibitem{Engelhardt:2018kcs}
N.~Engelhardt and A.~C. Wall, ``{Coarse Graining Holographic Black Holes},''
  \href{http://dx.doi.org/10.1007/JHEP05(2019)160}{{\em JHEP} {\bfseries 05}
  (2019) 160}, \href{http://arxiv.org/abs/1806.01281}{{\ttfamily
  arXiv:1806.01281 [hep-th]}}.

\bibitem{Nomura:2018aus}
Y.~Nomura and G.~N. Remmen, ``{Area Law Unification and the Holographic Event
  Horizon},'' \href{http://dx.doi.org/10.1007/JHEP08(2018)063}{{\em JHEP}
  {\bfseries 08} (2018) 063}, \href{http://arxiv.org/abs/1805.09339}{{\ttfamily
  arXiv:1805.09339 [hep-th]}}.

\bibitem{Bousso:2018fou}
R.~Bousso, Y.~Nomura, and G.~N. Remmen, ``{Outer Entropy and Quasilocal
  Energy},'' \href{http://dx.doi.org/10.1103/PhysRevD.99.046002}{{\em Phys.
  Rev. D} {\bfseries 99} (2019) 046002},
  \href{http://arxiv.org/abs/1812.06987}{{\ttfamily arXiv:1812.06987
  [hep-th]}}.

\bibitem{Bao:2018fso}
N.~Bao, A.~Chatwin-Davies, and G.~N. Remmen, ``{Entanglement of Purification
  and Multiboundary Wormhole Geometries},''
  \href{http://dx.doi.org/10.1007/JHEP02(2019)110}{{\em JHEP} {\bfseries 02}
  (2019) 110}, \href{http://arxiv.org/abs/1811.01983}{{\ttfamily
  arXiv:1811.01983 [hep-th]}}.

\bibitem{Freedman:2016zud}
M.~Freedman and M.~Headrick, ``{Bit Threads and Holographic Entanglement},''
  \href{http://dx.doi.org/10.1007/s00220-016-2796-3}{{\em Commun. Math. Phys.}
  {\bfseries 352} (2017) 407},
  \href{http://arxiv.org/abs/1604.00354}{{\ttfamily arXiv:1604.00354
  [hep-th]}}.

\bibitem{Bao:2019wcf}
N.~Bao, A.~Chatwin-Davies, J.~Pollack, and G.~N. Remmen, ``{Towards a Bit
  Threads Derivation of Holographic Entanglement of Purification},''
  \href{http://dx.doi.org/10.1007/JHEP07(2019)152}{{\em JHEP} {\bfseries 07}
  (2019) 152}, \href{http://arxiv.org/abs/1905.04317}{{\ttfamily
  arXiv:1905.04317 [hep-th]}}.

\bibitem{Bao:2017thr}
N.~Bao and G.~N. Remmen, ``{Bulk Connectedness and Boundary Entanglement},''
  \href{http://dx.doi.org/10.1209/0295-5075/121/60007}{{\em EPL} {\bfseries
  121} (2018) 60007}, \href{http://arxiv.org/abs/1703.00018}{{\ttfamily
  arXiv:1703.00018 [hep-th]}}.

\bibitem{Bao:2021vyq}
N.~Bao, A.~Chatwin-Davies, and G.~N. Remmen, ``{Entanglement Wedge Cross
  Section Inequalities from Replicated Geometries},''
  \href{http://dx.doi.org/10.1007/JHEP07(2021)113}{{\em JHEP} {\bfseries 07}
  (2021) 113}, \href{http://arxiv.org/abs/2106.02640}{{\ttfamily
  arXiv:2106.02640 [hep-th]}}.

\bibitem{Bao:2021ebo}
N.~Bao, J.~Harper, and G.~N. Remmen, ``{Holevo Information of Black Hole
  Mesostates},'' \href{http://dx.doi.org/10.1103/PhysRevD.105.026010}{{\em
  Phys. Rev. D} {\bfseries 105} (2022) 026010},
  \href{http://arxiv.org/abs/2103.06888}{{\ttfamily arXiv:2103.06888
  [hep-th]}}.

\bibitem{Chatwin-Davies:2023ofu}
A.~Chatwin-Davies, P.~Leung, and G.~N. Remmen, ``{Holographic Screen
  Sequestration},'' \href{http://dx.doi.org/10.1103/PhysRevD.109.046003}{{\em
  Phys. Rev. D} {\bfseries 109} (2024) 046003},
  \href{http://arxiv.org/abs/2312.06750}{{\ttfamily arXiv:2312.06750
  [hep-th]}}.

\bibitem{Hayden_2013}
P.~Hayden, M.~Headrick, and A.~Maloney, ``{Holographic Mutual Information is
  Monogamous},'' \href{http://dx.doi.org/10.1103/PhysRevD.87.046003}{{\em Phys.
  Rev. D} {\bfseries 87} (2013) 046003},
  \href{http://arxiv.org/abs/1107.2940}{{\ttfamily arXiv:1107.2940 [hep-th]}}.

\bibitem{Bao_2015}
N.~Bao, S.~Nezami, H.~Ooguri, B.~Stoica, J.~Sully, and M.~Walter, ``{The
  Holographic Entropy Cone},''
  \href{http://dx.doi.org/10.1007/JHEP09(2015)130}{{\em JHEP} {\bfseries 09}
  (2015) 130}, \href{http://arxiv.org/abs/1505.07839}{{\ttfamily
  arXiv:1505.07839 [hep-th]}}.

\bibitem{Ryu_2006}
S.~Ryu and T.~Takayanagi, ``{Holographic Derivation of Entanglement Entropy
  from AdS/CFT},'' \href{http://dx.doi.org/10.1103/PhysRevLett.96.181602}{{\em
  Phys. Rev. Lett.} {\bfseries 96} (2006) 181602},
  \href{http://arxiv.org/abs/hep-th/0603001}{{\ttfamily arXiv:hep-th/0603001}}.

\bibitem{Hubeny:2007xt}
V.~E. Hubeny, M.~Rangamani, and T.~Takayanagi, ``{A Covariant Holographic
  Entanglement Entropy Proposal},''
  \href{http://dx.doi.org/10.1088/1126-6708/2007/07/062}{{\em JHEP} {\bfseries
  07} (2007) 062}, \href{http://arxiv.org/abs/0705.0016}{{\ttfamily
  arXiv:0705.0016 [hep-th]}}.

\bibitem{Garfinkle:1990eq}
D.~Garfinkle and A.~Strominger, ``{Semiclassical Wheeler Wormhole
  Production},'' \href{http://dx.doi.org/10.1016/0370-2693(91)90665-D}{{\em
  Phys. Lett. B} {\bfseries 256} (1991) 146}.

\bibitem{Garfinkle:1993xk}
D.~Garfinkle, S.~B. Giddings, and A.~Strominger, ``{Entropy in Black Hole Pair
  Production},'' \href{http://dx.doi.org/10.1103/PhysRevD.49.958}{{\em Phys.
  Rev. D} {\bfseries 49} (1994) 958},
  \href{http://arxiv.org/abs/gr-qc/9306023}{{\ttfamily arXiv:gr-qc/9306023}}.

\bibitem{forthcoming}
D.~Gupta, M.~Headrick, and M.~Sasieta, {\em {\rm to appear}}.

\bibitem{Susskind_1993}
L.~Susskind, L.~Thorlacius, and J.~Uglum, ``{The Stretched Horizon and Black
  Hole Complementarity},''
  \href{http://dx.doi.org/10.1103/PhysRevD.48.3743}{{\em Phys. Rev. D}
  {\bfseries 48} (1993) 3743},
  \href{http://arxiv.org/abs/hep-th/9306069}{{\ttfamily arXiv:hep-th/9306069}}.

\bibitem{Almheiri_2013}
A.~Almheiri, D.~Marolf, J.~Polchinski, and J.~Sully, ``{Black Holes:
  Complementarity or Firewalls?},''
  \href{http://dx.doi.org/10.1007/JHEP02(2013)062}{{\em JHEP} {\bfseries 02}
  (2013) 062}, \href{http://arxiv.org/abs/1207.3123}{{\ttfamily arXiv:1207.3123
  [hep-th]}}.

\bibitem{Bao:2017who}
N.~Bao, S.~M. Carroll, A.~Chatwin-Davies, J.~Pollack, and G.~N. Remmen,
  ``{Branches of the Black Hole Wave Function Need Not Contain Firewalls},''
  \href{http://dx.doi.org/10.1103/PhysRevD.97.126014}{{\em Phys. Rev. D}
  {\bfseries 97} (2018) 126014},
  \href{http://arxiv.org/abs/1712.04955}{{\ttfamily arXiv:1712.04955
  [hep-th]}}.

\bibitem{Bousso:2012as}
R.~Bousso, ``{Complementarity Is Not Enough},''
  \href{http://dx.doi.org/10.1103/PhysRevD.87.124023}{{\em Phys. Rev. D}
  {\bfseries 87} (2013) 124023},
  \href{http://arxiv.org/abs/1207.5192}{{\ttfamily arXiv:1207.5192 [hep-th]}}.

\bibitem{Ellis:1973yv}
H.~G. Ellis, ``{Ether Flow Through a Drainhole: A Particle Model in General
  Relativity},'' \href{http://dx.doi.org/10.1063/1.1666161}{{\em J. Math.
  Phys.} {\bfseries 14} (1973) 104}.

\bibitem{Bronnikov:1973fh}
K.~A. Bronnikov, ``{Scalar-Tensor Theory and Scalar Charge},'' {\em
  \href{https://www.actaphys.uj.edu.pl/R/4/3/251/pdf}{Acta Phys. Polon. B}}
  {\bfseries 4} (1973) 251.

\bibitem{10.1119/1.15620}
M.~S. Morris and K.~S. Thorne, ``{Wormholes in Spacetime and their Use for
  Interstellar Travel: A Tool for Teaching General Relativity},''
  \href{http://dx.doi.org/10.1119/1.15620}{{\em Am. J. Phys.} {\bfseries 56}
  (1988) 395}.

\bibitem{Gao_2017}
P.~Gao, D.~L. Jafferis, and A.~C. Wall, ``{Traversable Wormholes via a Double
  Trace Deformation},'' \href{http://dx.doi.org/10.1007/JHEP12(2017)151}{{\em
  JHEP} {\bfseries 12} (2017) 151},
  \href{http://arxiv.org/abs/1608.05687}{{\ttfamily arXiv:1608.05687
  [hep-th]}}.

\bibitem{Maldacena_2017}
J.~Maldacena, D.~Stanford, and Z.~Yang, ``{Diving into Traversable
  Wormholes},'' \href{http://dx.doi.org/10.1002/prop.201700034}{{\em Fortsch.
  Phys.} {\bfseries 65} (2017) 1700034},
  \href{http://arxiv.org/abs/1704.05333}{{\ttfamily arXiv:1704.05333
  [hep-th]}}.

\bibitem{Banados:1992wn}
M.~Ba\~nados, C.~Teitelboim, and J.~Zanelli, ``{Black Hole in Three-Dimensional
  Spacetime},'' \href{http://dx.doi.org/10.1103/PhysRevLett.69.1849}{{\em Phys.
  Rev. Lett.} {\bfseries 69} (1992) 1849},
  \href{http://arxiv.org/abs/hep-th/9204099}{{\ttfamily arXiv:hep-th/9204099}}.

\bibitem{Shenker:2013pqa}
S.~H. Shenker and D.~Stanford, ``{Black Holes and the Butterfly Effect},''
  \href{http://dx.doi.org/10.1007/JHEP03(2014)067}{{\em JHEP} {\bfseries 03}
  (2014) 067}, \href{http://arxiv.org/abs/1306.0622}{{\ttfamily arXiv:1306.0622
  [hep-th]}}.

\bibitem{Bao_2018}
N.~Bao, A.~Chatwin-Davies, J.~Pollack, and G.~N. Remmen, ``{Traversable
  Wormholes as Quantum Channels: Exploring CFT Entanglement Structure and
  Channel Capacity in Holography},''
  \href{http://dx.doi.org/10.1007/JHEP11(2018)071}{{\em JHEP} {\bfseries 11}
  (2018) 071}, \href{http://arxiv.org/abs/1808.05963}{{\ttfamily
  arXiv:1808.05963 [hep-th]}}.

\bibitem{Graham:2007va}
N.~Graham and K.~D. Olum, ``{Achronal Averaged Null Energy Condition},''
  \href{http://dx.doi.org/10.1103/PhysRevD.76.064001}{{\em Phys. Rev. D}
  {\bfseries 76} (2007) 064001},
  \href{http://arxiv.org/abs/0705.3193}{{\ttfamily arXiv:0705.3193 [gr-qc]}}.

\bibitem{Wald:1991xn}
R.~M. Wald and U.~Yurtsever, ``{General Proof of the Averaged Null Energy
  Condition for a Massless Scalar Field in Two-Dimensional Curved Spacetime},''
  \href{http://dx.doi.org/10.1103/PhysRevD.44.403}{{\em Phys. Rev. D}
  {\bfseries 44} (1991) 403}.

\bibitem{Visser:1994jb}
M.~Visser, ``{Scale Anomalies Imply Violation of the Averaged Null Energy
  Condition},'' \href{http://dx.doi.org/10.1016/0370-2693(95)00303-3}{{\em
  Phys. Lett. B} {\bfseries 349} (1995) 443},
  \href{http://arxiv.org/abs/gr-qc/9409043}{{\ttfamily arXiv:gr-qc/9409043}}.

\bibitem{Urban:2009yt}
D.~Urban and K.~D. Olum, ``{Averaged Null Energy Condition Violation in a
  Conformally Flat Spacetime},''
  \href{http://dx.doi.org/10.1103/PhysRevD.81.024039}{{\em Phys. Rev. D}
  {\bfseries 81} (2010) 024039},
  \href{http://arxiv.org/abs/0910.5925}{{\ttfamily arXiv:0910.5925 [gr-qc]}}.

\bibitem{Hartman:2023qdn}
T.~Hartman and G.~Mathys, ``{Averaged Null Energy and the Renormalization
  Group},'' \href{http://dx.doi.org/10.1007/JHEP12(2023)139}{{\em JHEP}
  {\bfseries 12} (2023) 139}, \href{http://arxiv.org/abs/2309.14409}{{\ttfamily
  arXiv:2309.14409 [hep-th]}}.

\bibitem{Maldacena:2018gjk}
J.~Maldacena, A.~Milekhin, and F.~Popov, ``{Traversable Wormholes in Four
  Dimensions},'' \href{http://dx.doi.org/10.1088/1361-6382/acde30}{{\em Class.
  Quant. Grav.} {\bfseries 40} (2023) 155016},
  \href{http://arxiv.org/abs/1807.04726}{{\ttfamily arXiv:1807.04726
  [hep-th]}}.

\bibitem{tHooft:1984kcu}
G.~'t~Hooft, ``{On the Quantum Structure of a Black Hole},''
  \href{http://dx.doi.org/10.1016/0550-3213(85)90418-3}{{\em Nucl. Phys. B}
  {\bfseries 256} (1985) 727}.

\bibitem{MembraneParadigm}
K.~S. Thorne, R.~Price, and D.~Macdonald, {\em {Black Holes: The Membrane
  Paradigm}}.
\newblock Yale University Press,
1986.
\newblock
%%CITATION = ISBN-9780300037708 ETC.;%%.

\bibitem{SusskindHolography}
L.~Susskind, ``{The World as a Hologram},''
  \href{http://dx.doi.org/10.1063/1.531249}{{\em J.Math.Phys.} {\bfseries 36}
  (1995) 6377},
\href{http://arxiv.org/abs/hep-th/9409089}{{\ttfamily arXiv:hep-th/9409089
  [hep-th]}}.
%%CITATION = HEP-TH/9409089;%%.

\bibitem{tHooft:1993dmi}
G.~'t~Hooft, ``{Dimensional Reduction in Quantum Gravity},'' {\em Conf. Proc.
  C} {\bfseries 930308} (1993) 284,
  \href{http://arxiv.org/abs/gr-qc/9310026}{{\ttfamily arXiv:gr-qc/9310026}}.

\bibitem{Bousso:2002ju}
R.~Bousso, ``{The Holographic Principle},''
  \href{http://dx.doi.org/10.1103/RevModPhys.74.825}{{\em Rev. Mod. Phys.}
  {\bfseries 74} (2002) 825},
  \href{http://arxiv.org/abs/hep-th/0203101}{{\ttfamily arXiv:hep-th/0203101}}.

\bibitem{AdSCFT}
J.~M. Maldacena, ``{The Large-$N$ Limit of Superconformal Field Theories and
  Supergravity},'' \href{http://dx.doi.org/10.1023/A:1026654312961}{{\em
  Int.J.Theor.Phys.} {\bfseries 38} (1999) 1113},
\href{http://arxiv.org/abs/hep-th/9711200}{{\ttfamily arXiv:hep-th/9711200
  [hep-th]}}.
%%CITATION = HEP-TH/9711200;%%.

\bibitem{MAGOO}
O.~Aharony, S.~S. Gubser, J.~M. Maldacena, H.~Ooguri, and Y.~Oz, ``{Large-$N$
  Field Theories, String Theory and Gravity},''
  \href{http://dx.doi.org/10.1016/S0370-1573(99)00083-6}{{\em Phys.Rept.}
  {\bfseries 323} (2000) 183},
\href{http://arxiv.org/abs/hep-th/9905111}{{\ttfamily arXiv:hep-th/9905111
  [hep-th]}}.
%%CITATION = HEP-TH/9905111;%%.

\bibitem{Witten:1998qj}
E.~Witten, ``{Anti-de~Sitter Space and Holography},'' {\em Adv. Theor. Math.
  Phys.} {\bfseries 2} (1998) 253,
\href{http://arxiv.org/abs/hep-th/9802150}{{\ttfamily arXiv:hep-th/9802150
  [hep-th]}}.
%%CITATION = HEP-TH/9802150;%%.

\bibitem{SusskindWitten}
L.~Susskind and E.~Witten, ``{The Holographic Bound in Anti-de~Sitter Space},''
\href{http://arxiv.org/abs/hep-th/9805114}{{\ttfamily arXiv:hep-th/9805114
  [hep-th]}}.
%%CITATION = HEP-TH/9805114;%%.

\bibitem{penington2020entanglementwedgereconstructioninformation}
G.~Penington, ``{Entanglement Wedge Reconstruction and the Information
  Paradox},'' \href{http://dx.doi.org/10.1007/JHEP09(2020)002}{{\em JHEP}
  {\bfseries 09} (2020) 002}, \href{http://arxiv.org/abs/1905.08255}{{\ttfamily
  arXiv:1905.08255 [hep-th]}}.

\bibitem{Almheiri_2019}
A.~Almheiri, N.~Engelhardt, D.~Marolf, and H.~Maxfield, ``{The Entropy of Bulk
  Quantum Fields and the Entanglement Wedge of an Evaporating Black Hole},''
  \href{http://dx.doi.org/10.1007/JHEP12(2019)063}{{\em JHEP} {\bfseries 12}
  (2019) 063}, \href{http://arxiv.org/abs/1905.08762}{{\ttfamily
  arXiv:1905.08762 [hep-th]}}.

\bibitem{Geng:2020qvw}
H.~Geng and A.~Karch, ``{Massive islands},''
  \href{http://dx.doi.org/10.1007/JHEP09(2020)121}{{\em JHEP} {\bfseries 09}
  (2020) 121}, \href{http://arxiv.org/abs/2006.02438}{{\ttfamily
  arXiv:2006.02438 [hep-th]}}.

\bibitem{Geng:2020fxl}
H.~Geng, A.~Karch, C.~Perez-Pardavila, S.~Raju, L.~Randall, M.~Riojas, and
  S.~Shashi, ``{Information Transfer with a Gravitating Bath},''
  \href{http://dx.doi.org/10.21468/SciPostPhys.10.5.103}{{\em SciPost Phys.}
  {\bfseries 10} no.~5, (2021) 103},
  \href{http://arxiv.org/abs/2012.04671}{{\ttfamily arXiv:2012.04671
  [hep-th]}}.

\bibitem{Geng:2021hlu}
H.~Geng, A.~Karch, C.~Perez-Pardavila, S.~Raju, L.~Randall, M.~Riojas, and
  S.~Shashi, ``{Inconsistency of islands in theories with long-range
  gravity},'' \href{http://dx.doi.org/10.1007/JHEP01(2022)182}{{\em JHEP}
  {\bfseries 01} (2022) 182}, \href{http://arxiv.org/abs/2107.03390}{{\ttfamily
  arXiv:2107.03390 [hep-th]}}.

\bibitem{Raju:2021lwh}
S.~Raju, ``{Failure of the split property in gravity and the information
  paradox},'' \href{http://dx.doi.org/10.1088/1361-6382/ac482b}{{\em Class.
  Quant. Grav.} {\bfseries 39} (2022) 064002},
  \href{http://arxiv.org/abs/2110.05470}{{\ttfamily arXiv:2110.05470
  [hep-th]}}.

\bibitem{Geng:2021wcq}
H.~Geng, Y.~Nomura, and H.-Y. Sun, ``{Information paradox and its resolution in
  de Sitter holography},''
  \href{http://dx.doi.org/10.1103/PhysRevD.103.126004}{{\em Phys. Rev. D}
  {\bfseries 103} (2021) 126004},
  \href{http://arxiv.org/abs/2103.07477}{{\ttfamily arXiv:2103.07477
  [hep-th]}}.

\bibitem{Bousso:2023kdj}
R.~Bousso and G.~Penington, ``{Islands far outside the horizon},''
  \href{http://dx.doi.org/10.1007/JHEP11(2024)164}{{\em JHEP} {\bfseries 11}
  (2024) 164}, \href{http://arxiv.org/abs/2312.03078}{{\ttfamily
  arXiv:2312.03078 [hep-th]}}.

\bibitem{Antonini:2025sur}
S.~Antonini, C.-H. Chen, H.~Maxfield, and G.~Penington, ``{An apologia for
  islands},'' \href{http://arxiv.org/abs/2506.04311}{{\ttfamily
  arXiv:2506.04311 [hep-th]}}.

\bibitem{Okada_2024}
M.~Okada and Y.~Tachikawa, ``{Noninvertible Symmetries Act Locally by Quantum
  Operations},'' \href{http://dx.doi.org/10.1103/PhysRevLett.133.191602}{{\em
  Phys. Rev. Lett.} {\bfseries 133} (2024) 191602},
  \href{http://arxiv.org/abs/2403.20062}{{\ttfamily arXiv:2403.20062
  [hep-th]}}.

\bibitem{bennett2002entanglementassistedcapacityquantumchannel}
C.~H. Bennett, P.~W. Shor, J.~A. Smolin, and A.~V. Thapliyal,
  ``{Entanglement-Assisted Capacity of a Quantum Channel and the Reverse
  Shannon Theorem},'' \href{http://dx.doi.org/10.1109/TIT.2002.802612}{{\em
  IEEE Trans. Info. Theor.} {\bfseries 48} (2002) 2637}.

\bibitem{May:2019yxi}
A.~May, ``{Quantum Tasks in Holography},''
  \href{http://dx.doi.org/10.1007/JHEP10(2019)233}{{\em JHEP} {\bfseries 10}
  (2019) 233}, \href{http://arxiv.org/abs/1902.06845}{{\ttfamily
  arXiv:1902.06845 [hep-th]}}.
  \href{https://doi.org/10.1007/JHEP01(2020)080}{[Erratum: {\it JHEP} {\bf 01}
  (2020) 080]}.

\bibitem{cmp/1103758945}
S.~Schlieder, ``{Some Remarks about the Localization of States in a Quantum
  Field Theory},'' \href{http://dx.doi.org/10.1007/BF01645904}{{\em Comm. Math.
  Phys.} {\bfseries 1} (1965) 265}.

\bibitem{Bekenstein:1973ur}
J.~D. Bekenstein, ``{Black Holes and Entropy},''
  \href{http://dx.doi.org/10.1103/PhysRevD.7.2333}{{\em Phys. Rev. D}
  {\bfseries 7} (1973) 2333}.

\bibitem{HawkingParticle}
S.~W. Hawking, ``{Particle Creation by Black Holes},''
  \href{http://dx.doi.org/10.1007/BF02345020}{{\em Commun. Math. Phys.}
  {\bfseries 43} (1975) 199}.
  \href{https://doi.org/10.1007/BF01608497}{[Erratum: {\it Commun. Math. Phys.}
  {\bf 46} (1976) 206]}.

\bibitem{strohmaier2023timeliketubetheoremcurved}
A.~Strohmaier and E.~Witten, ``{The Timelike Tube Theorem in Curved
  Spacetime},'' \href{http://dx.doi.org/10.1007/s00220-024-05009-3}{{\em
  Commun. Math. Phys.} {\bfseries 405} (2024) 153},
  \href{http://arxiv.org/abs/2303.16380}{{\ttfamily arXiv:2303.16380
  [hep-th]}}.

\bibitem{Chandrasekaran_2023}
V.~Chandrasekaran, R.~Longo, G.~Penington, and E.~Witten, ``{An Algebra of
  Observables for de~Sitter Space},''
  \href{http://dx.doi.org/10.1007/JHEP02(2023)082}{{\em JHEP} {\bfseries 02}
  (2023) 082}, \href{http://arxiv.org/abs/2206.10780}{{\ttfamily
  arXiv:2206.10780 [hep-th]}}.

\bibitem{Zanardi_2004}
P.~Zanardi, D.~A. Lidar, and S.~Lloyd, ``{Quantum Tensor Product Structures are
  Observable Induced},''
  \href{http://dx.doi.org/10.1103/PhysRevLett.92.060402}{{\em Phys. Rev. Lett.}
  {\bfseries 92} (2004) 060402},
  \href{http://arxiv.org/abs/quant-ph/0308043}{{\ttfamily
  arXiv:quant-ph/0308043}}.

\bibitem{cmp/1103858407}
E.~H. Lieb and D.~W. Robinson, ``{The Finite Group Velocity of Quantum Spin
  Systems},'' \href{http://dx.doi.org/10.1007/BF01645779}{{\em Commun. Math.
  Phys.} {\bfseries 28} (1972) 251}.

\bibitem{Roberts_2016}
D.~A. Roberts and B.~Swingle, ``{Lieb-Robinson Bound and the Butterfly Effect
  in Quantum Field Theories},''
  \href{http://dx.doi.org/10.1103/PhysRevLett.117.091602}{{\em Phys. Rev.
  Lett.} {\bfseries 117} (2016) 091602},
  \href{http://arxiv.org/abs/1603.09298}{{\ttfamily arXiv:1603.09298
  [hep-th]}}.

\bibitem{Roberts_2014}
D.~A. Roberts, D.~Stanford, and L.~Susskind, ``{Localized Shocks},''
  \href{http://dx.doi.org/10.1007/JHEP03(2015)051}{{\em JHEP} {\bfseries 03}
  (2015) 051}, \href{http://arxiv.org/abs/1409.8180}{{\ttfamily arXiv:1409.8180
  [hep-th]}}.

\end{thebibliography}\endgroup
\end{document}